\documentclass[aps,prb,twocolumn,10pt]{revtex4-1}

\usepackage{amsmath}
\usepackage{amssymb}
\usepackage{graphicx}
\usepackage{multirow}
\newcommand{\qsgw}{QS\emph{GW}}
\usepackage[usenames,dvipsnames]{color}

\begin{document}

\title{Electronic structure and magnetic properties of Gd-doped and Eu-rich EuO}
\author{J. M. An and K. D. Belashchenko}

\affiliation{Department of Physics and Astronomy and Nebraska Center for Materials and Nanoscience, University of Nebraska-Lincoln, Lincoln, Nebraska 68588, USA}
\date{\today}

\begin{abstract}
The effects of Gd doping and O vacancies on the magnetic interaction and Curie temperature $T_C$ of EuO are studied using first-principles calculations. Linear response calculations in the virtual crystal approximation show a broad maximum in the Curie temperature as a function of doping, which results from the combination of the saturating contribution from indirect exchange and a decreasing contribution from the $f$-$d$ hopping mechanism. Non-Heisenberg interaction at low doping levels and its effect on $T_C$ are examined. The electronic structure of a substitutional Gd and of an O vacancy in EuO are evaluated. When the $4f$ spins are disordered, the impurity state goes from single to double occupation, but correlated bound magnetic polarons are not ruled out. At higher vacancy concentrations typical for Eu-rich EuO films, the impurity states broaden into bands and remain partially filled. To go beyond the homogeneous doping picture, magnetostructural cluster expansions are constructed, which describe the modified exchange parameters near Gd dopants or O vacancies. Thermodynamic properties are studied using Monte Carlo simulations. The Curie temperature for Gd-doped EuO agrees with the results of the virtual crystal approximation and shows a maximum of about 150 K. At 3.125\% vacancy concentration the $T_C$ increases to 120 K, consistent with experimental data for Eu-rich film samples.
\end{abstract}

\maketitle

\section{Introduction}

EuO is a ferromagnetic oxide semiconductor \cite{Matthias,Wachter} with unusual magnetic and transport properties.\cite{Kasuya-RMP,Nagaev,Mauger-Godart} Electron doping by trivalent rare-earth elements nearly doubles its Curie temperature $T_C$, while the $M(T)$ curve deviates significantly from the Brillouin function.\cite{Shafer,Mauger-Godart,Schmehl,Sutarto,Mairoser} A similar enhancement of $T_C$ is observed in Eu-rich EuO films.\cite{Massenet,Borukhovich,Matsumoto,Barbagallo} The magnetic transition in bulk Eu-rich EuO is accompanied by a metal-insulator transition,\cite{Oliver,Oliver2,Petrich} while the $T_C$ is not enhanced. Recent interest in EuO is due to its potential applications in spintronics as a spin filter in tunnel junctions \cite{Moodera} in insulating pure form and as a high spin-polarization material \cite{Steeneken} when doped. It was shown that EuO can be grown epitaxially on various substrates including Si, GaN, and GaAs.\cite{Schmehl,Swartz}

Ferromagnetism in pure EuO is due to the $4f$-$5d$ hopping mechanism.\cite{Kasuya-IBM} First-principles calculations confirm that magnetic interaction is short-range and ferromagnetic.\cite{Larson,Kunes,Savrasov} Large enhancement of $T_C$ under doping is induced by the filling of the conduction band and the associated indirect exchange mechanism. This mechanism is physically of the Ruderman-Kittel-Kasuya-Yosida (RKKY) type, \cite{RKKY} but at low doping levels it deviates from the Heisenberg form owing to the half-metallic character of the conduction band in the ferromagnetic state.\cite{Mauger,Nagaev,Takahashi} This non-Heisenberg character may be responsible for the peculiar shoulder observed in the $M(T)$ curves for Gd-doped EuO.

$T_C$ goes through a broad maximum as a function of Gd concentration.\cite{Sutarto} First-principles calculations have been limited to the virtual crystal approximation and were based on the fitting of the total energies for several magnetic configurations to the Heisenberg model.\cite{Ingle,Savrasov} These calculations yield a broad maximum in $T_C$ in agreement with experiment. It was, however, argued \cite{Takahashi} that this maximum is associated with dynamical effects. On the other hand, Mairoser \emph{et al.}\cite{Mairoser} observed that the carrier density appears to fall behind the actual Gd concentration and raised the question whether the intrinsic limit of $T_C$ has been achieved.

The metal-insulator transition in Eu-rich EuO involves impurity levels on O vacancies,\cite{Oliver,Oliver2,Torrance,Laks,Sinjukow} which absorb free charge carriers from the conduction band in the paramagnetic state. Whether a similar transition exists in Gd-doped EuO \cite{Godart,Samokhvalov,Arnold,Shen} due to shallow donor levels is not fully clear, as the transition is not observed in all measurements and may be due to the presence of oxygen vacancies.\cite{Schoenes}

Here we investigate the electronic structure and exchange interaction in Gd-doped and Eu-rich EuO using first-principles electronic structure methods combined with both linear response and total energy calculations.
The paper is organized as follows. Linear response calculations in the virtual crystal approximation are described in Section \ref{VCA}. The electronic structure of isolated Gd dopants and O vacancies, and in particular the impurity levels introduced by O vacancies, are discussed in Section \ref{Isolated}. The magnetostructural cluster expansions for Gd-doped and O-deficient EuO are constructed in Section \ref{sec:MSCE}. Thermodynamic properties are evaluated using Monte Carlo simulations in Section \ref{sec:MC}. The conclusions are drawn in Section \ref{Conclusions}. The results of Sections \ref{sec:MSCE} and \ref{sec:MC} for Gd-doped EuO were, in part, briefly reported in conference proceedings.\cite{SPIE}

\section{Exchange interaction in the virtual crystal approximation}\label{VCA}

In this section we evaluate the effect of electron doping on the exchange interaction and Curie temperature of EuO using the linear response method for the calculation of exchange parameters \cite{Jij-JMMM} within the density functional theory (DFT).
We use Green's function-based formulation of the tight-binding linear muffin-tin orbital (TB-LMTO) method \cite{Andersen,Turek} in the atomic sphere approximation (ASA). The lattice constant is fixed at the experimental value 5.14 \AA. To improve agreement with the full-potential band structure, two empty spheres (ES) per formula unit are placed at the centers of all nearest-neighbor Eu tetrahedra. The electron doping is introduced within the virtual crystal approximation (VCA) by adding the electron doping concentration $x$ to the nuclear charge of the Eu atom.

\begin{figure}[hbt]
\includegraphics[width=0.45\textwidth,angle=0,clip]{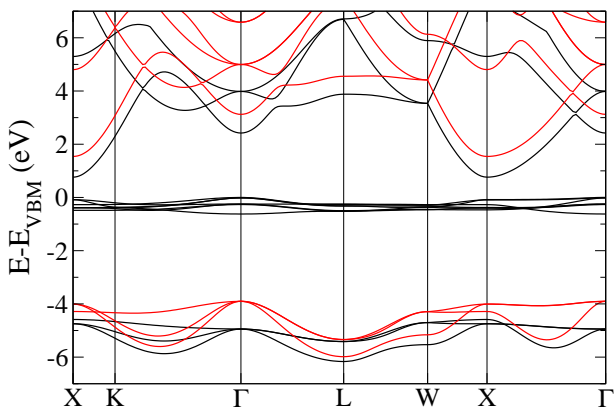}
\caption{(Color online) EuO band structure corresponding to the LMTO Model 1. Black and red (gray) lines show majority and minority-spin bands, respectively.}
\label{bandEuO}
\end{figure}

In order to expose the different mechanisms of exchange interaction, we have considered three different models. Model 1 is designed to provide the best fit to the quasiparticle self-consistent $GW$ (\qsgw) calculation.\cite{An-PRB} The basis set includes $6s$, $6p$, $5d$, and $4f$ states on Eu, $3s$, $2p$, and $3d$ states on O, and $1s$, $2s$, and $3d$ states on the empty spheres. The LMTO sphere radii were chosen as 1.79 \AA\ for Eu, 1.16 \AA\ for O, and 0.74 \AA\ for the empty spheres. Strong correlations within the $4f$ shell were included using the LDA$+U$ method \cite{prb-52-r5467} with $U=6.12$ eV and $J=0.6$ eV. In addition, an auxiliary external potential of $V_{2p}=-5$ eV was applied to the O $2p$ states in order to place them at the correct binding energy.\cite{An-PRB,Eastman} The resulting band structure of undoped EuO is shown in Fig.\ \ref{bandEuO}. The indirect $\Gamma$-X band gap is 0.76 eV, and the exchange splitting of the conduction band minimum (CBM) at the X point is 0.78 eV. The band gap is a little too small (the experimental optical absorption gap is 0.95 eV at low temperatures\cite{Dimmock}), and the CBM splitting is somewhat larger than the experimentally reported 0.6 eV.\cite{Steeneken}

In models 2 and 3 the ``open core'' approximation is used for the Eu $4f$ electrons, whereby these states are treated as a polarized core and thereby not allowed to hybridize. In these models the LMTO sphere radii were 1.69, 1.27, and 0.86 \AA\ for Eu, O, and empty spheres, respectively. In Model 2, a $-5$ eV external potential is applied to the O $2p$ states just as in Model 1, but in Model 3 this is not done. The difference between these two models highlights the effects of oxygen hybridization with the conduction-band states.

\begin{figure}[hbt]
\includegraphics[width=0.45\textwidth,angle=0,clip]{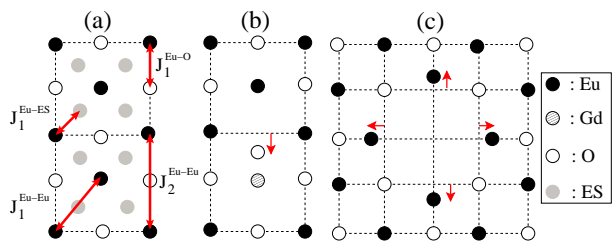}
\caption{(Color online) (a) Illustration of the pairs of atoms corresponding to the exchange parameters calculated in LMTO. (b) Shift of the O atom toward Gd in a Eu-O-Gd chain. (c) Shift of the Eu atoms away from an O vacancy. The empty spheres are shown in projection onto the plane of the figures. The shifts in panels (b) and (c) are exaggerated.}
\label{scheme}
\end{figure}

Table \ref{vcajij} lists the important exchange parameters (see Fig.\ \ref{scheme}a for an illustration) calculated for the three models at zero and 5\% electron doping levels, as well as the differences between them. First, let us comment on the physical meaning of these parameters. In the linear response method \cite{Jij-JMMM} the exchange parameter for a pair of sites $i$ and $j$ is defined as the second derivative of the total energy with respect to the angles of rotation of the exchange-correlation fields in the corresponding atomic spheres: $J_{ij}=\partial^2 E/\partial\theta_i\partial\theta_j$. In the LMTO setups all sites carry local moments, including O and the empty spheres ES. Therefore, one can formally define the pair exchange parameters for the Eu-O pairs, Eu-ES pairs, etc. Table \ref{vcajij} lists all such parameters that are not negligibly small in undoped EuO (others are less than 0.01 meV). In the doped case a long-range indirect interaction sets in, which will be discussed separately below. Although individual pair parameters for this long-range part are not listed in Table \ref{vcajij}, it contains the full lattice sum of the Eu-Eu pair parameters ($J^\textrm{Eu-Eu}_0$), as well as the full lattice sum of all pair parameters connecting a given site X to the rest of the lattice ($J^\textrm{X}_0$). Note that the exchange parameters are first calculated in reciprocal space and then Fourier-transformed to real space. The lattice sums are thus calculated exactly, without a real-space cutoff.

A sensible effective model of exchange interaction in EuO should only include Eu spins as dynamical variables. The small magnetic moments on O are induced by exchange interaction (and hybridization for Model 1) with the $4f$ states, while those on ES sites simply reflect the extension of the Eu and O orbitals beyond their atomic spheres. These local moments do not fluctuate independently, but should be assumed to adiabatically follow the fluctuating $4f$ moments on Eu. We should not, therefore, interpret the exchange parameters connecting O and ES to other sites as having direct physical meaning, but these interactions should contribute to the true effective Eu-Eu exchange parameters. For the estimate of the Curie temperature $T_C$, in the mean-field approximation we should consider the total Weiss field acting on a given Eu spin. In doing so, we should exclude the part attributable to the tails of the orbitals on the central Eu site itself. In other words, we should exclude the spurious effects of the orbital ``tails'' being fixed while its ``head'' is rotated, which is what happens, for example, when a Eu-O exchange parameter is evaluated using the linear response technique in LMTO. Since such parameters (Eu-O and Eu-ES) extend only to the nearest-neighbor sites of the given kind, on a crude level we can simply reduce each such parameter by a factor $(N_j-1)/N_j$, where $N_j$ is the number of nearest-neighbor Eu atoms for site $j$ (which is either O or ES). For Model 1 we can then estimate the ``true'' effective field parameter $J_0$ as $J^\textrm{Eu-Eu}_0+6(6/7)J^\textrm{Eu-O}_1+8(3/4)J^\textrm{Eu-ES}_1=18.1$ meV, which is only slightly smaller than $J^\textrm{Eu}_0=19.3$ meV. The classical mean-field estimate of $T_C=(2/3)J_0$ for undoped EuO then becomes 140 K (we use classical mean-field estimates of $T_C$ to facilitate comparison with classical Monte Carlo results in the following). This value is about twice as large compared to the experimental value. This overestimation can only partially be attributed to the mean-field approximation, and the remaining error is probably related to the ASA.

\begin{table}[hbt]
\caption{Exchange parameters (meV) for pure and electron-doped EuO in the virtual crystal approximation.}
\begin{tabular}{|l|c|c|c||c|c|c||c|c|c|}
\hline
\multicolumn{1}{|c|}{Model}   & \multicolumn{3}{c||}{\#1} & \multicolumn{3}{c||}{\#2} & \multicolumn{3}{c|}{\#3} \\
\hline
\multicolumn{1}{|c|}{Eu $4f$} & \multicolumn{3}{c||}{Valence} & \multicolumn{3}{c||}{Core} & \multicolumn{3}{c|}{Core} \\
\hline
\multicolumn{1}{|c|}{$V_{2p}$} & \multicolumn{3}{c||}{$-5$ eV} & \multicolumn{3}{c||}{$-5$ eV} & \multicolumn{3}{c|}{0} \\
\hline
\multicolumn{1}{|c|}{Doping}  & 0\% & 5\% & Diff & 0\% & 5\% & Diff & 0\% & 5\% & Diff \\
\hline
$J^\textrm{Eu-Eu}_1$ & 1.05 & 1.22 & 0.17 & 0.12 & 0.33 & 0.21 & 0.13 & 0.32 & 0.19 \\
$J^\textrm{Eu-Eu}_2$ & 0.22 & 0.30 & 0.08 & $-0.14$ & $-0.04$ & 0.10 & $-0.37$ & $-0.26$ & 0.10 \\
$J^\textrm{Eu-O}_1$  & 0.27 & 0.32 & 0.04 & 0.21 & 0.24 & 0.03 & 0.44 & 0.47 & 0.04 \\
$J^\textrm{Eu-ES}_1$ & 0.43 & 0.53 & 0.11 & 0.03 & 0.11 & 0.08 & 0.07 & 0.16 & 0.09 \\
$J^\textrm{Eu-Eu}_0$ & 14.1 & 18.6 & 4.5  & 0.63 & 6.0  & 5.4  & $-0.75$ & 4.1 & 4.9 \\
$J^\textrm{Eu}_0$    & 19.3 & 25.7 & 6.3  & 2.1  & 8.9  & 6.8  & 2.4  & 8.6 & 6.2 \\
$J^\textrm{O}_0$     & 1.73 & 2.01 & 0.29 & 1.29 & 1.52 & 0.22 & 2.78 & 3.05 & 0.27 \\
$J^\textrm{ES}_0$    & 1.84 & 2.72 & 0.88 & 0.12 & 0.78 & 0.65 & 0.28 & 0.95 & 0.66 \\
\hline
\end{tabular}
\label{vcajij}
\end{table}

Comparison between the three models provides information about the exchange mechanisms. In Model 2, where the $4f$ electrons are not allowed to hybridize, $J^\textrm{Eu-Eu}_1$ and $J^\textrm{Eu-ES}_1$ are reduced by an order of magnitude, while $J^\textrm{Eu-Eu}_2$ becomes negative.
The Eu-O coupling, on the other hand, is only weakly affected.
These features confirm that the dominant part of the ferromagnetic coupling in EuO is due to the hybridization of $4f$ states with the conduction-band Eu states. \cite{Kasuya-IBM}
Note that in Ref.\ \onlinecite{Kunes} strong antiferromagnetic coupling was found in the open-core approximation, the origin of which is unclear to us. Weak ferromagnetic coupling in our open-core calculations is due to the Bloembergen-Rowland mechanism,\cite{BR} which involves the hybridization of O states with the empty conduction band. The downward shift of the O $2p$ states in Model 2 reduces this mechanism compared to Model 3, but it is quite small in both Models 2 and 3.

Let us now focus on the effects of electron doping. It is seen from Table \ref{vcajij} that the changes in the exchange parameters due to 5\% electron doping are quite similar across all three models. The dominant effect of doping is the introduction of an additional indirect contribution to Eu-Eu coupling, while the parameters connecting Eu to O and ES sites are only weakly modified. This indirect-exchange enhancement of the Eu-Eu coupling is not sensitive to the position of the O $2p$ states (compare Models 2 and 3), as expected for this mechanism.

\begin{figure}[htb]
\includegraphics[width=0.45\textwidth,angle=0,clip]{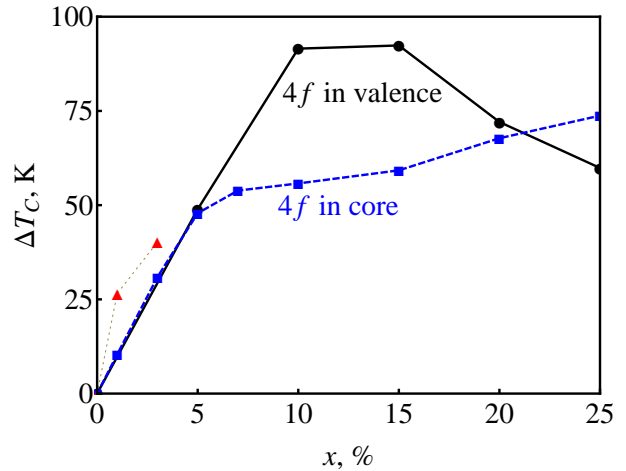}
\caption{(Color online) Enhancement of the Curie temperature ($\Delta T_C$) under electron doping for Model 1 (solid black) and Model 3 (dashed blue) calculated in the mean-field approximation. Red triangles: $\Delta T_C$ for Model 3 with a correction for non-Heisenberg interaction (see text).}
\label{tcdop}
\end{figure}

Fig.\ \ref{tcdop} shows the enhancement of $T_C$ in EuO as a function of electron doping $x$ evaluated in the mean-field approximation for Models 1 and 3 as $\Delta T_C=(2/3)(J^\mathrm{Eu}_0(x)-J^\mathrm{Eu}_0(0))$.
In Model 1 the $T_C$ increases with the doping concentration $x$ up to an optimal doping level of 10-15\% and then declines. This behavior agrees with earlier theoretical results \cite{Ingle,Savrasov} and with experimental data (see e.\ g.\ Ref.\ \onlinecite{Mairoser}). On the other hand, for Model 3 with $4f$ electrons treated as open core, $\Delta T_C$ increases monotonically as a function of doping.

Takahashi \cite{Takahashi2} argued that the maximum in the doping dependence of $T_C$ in EuO is due to dynamical effects in the effective interaction between $4f$ and conduction electrons. In his model $T_C$ saturates as a function of doping unless these dynamical effects are taken into account; he also suggested that the maximum in $T_C$ found in first-principles calculations of Refs.\ \onlinecite{Ingle,Savrasov} is due to the use of total energy calculations, which partially take into account the effects of spin disorder. However, our linear response calculations applied to the ferromagnetic state still produce a maximum in $T_C$, but only if the $4f$ electrons are included in the basis set. These results strongly suggest that the observed maximum in $T_C$ is due to the competition of two exchange mechanisms. The contribution of indirect exchange mediated by conduction electrons increases monotonically with doping, but the rate of its increase drops near the point where the Fermi level reaches the bottom of the minority-spin conduction band. On the other hand, the $f$-$d$ hopping mechanism, which dominates in pure EuO, is suppressed by the filling of the conduction band and by the loss of its spin polarization. The combination of these two effects leads to a broad maximum in $T_C$. As we mentioned above, the strength of the $f$-$d$ hopping mechanism is overestimated in our calculations, so one may expect that the decline of $T_C$ may also be somewhat overestimated.

\begin{figure}[htb]
\includegraphics[height=0.35\textwidth,angle=0,clip]{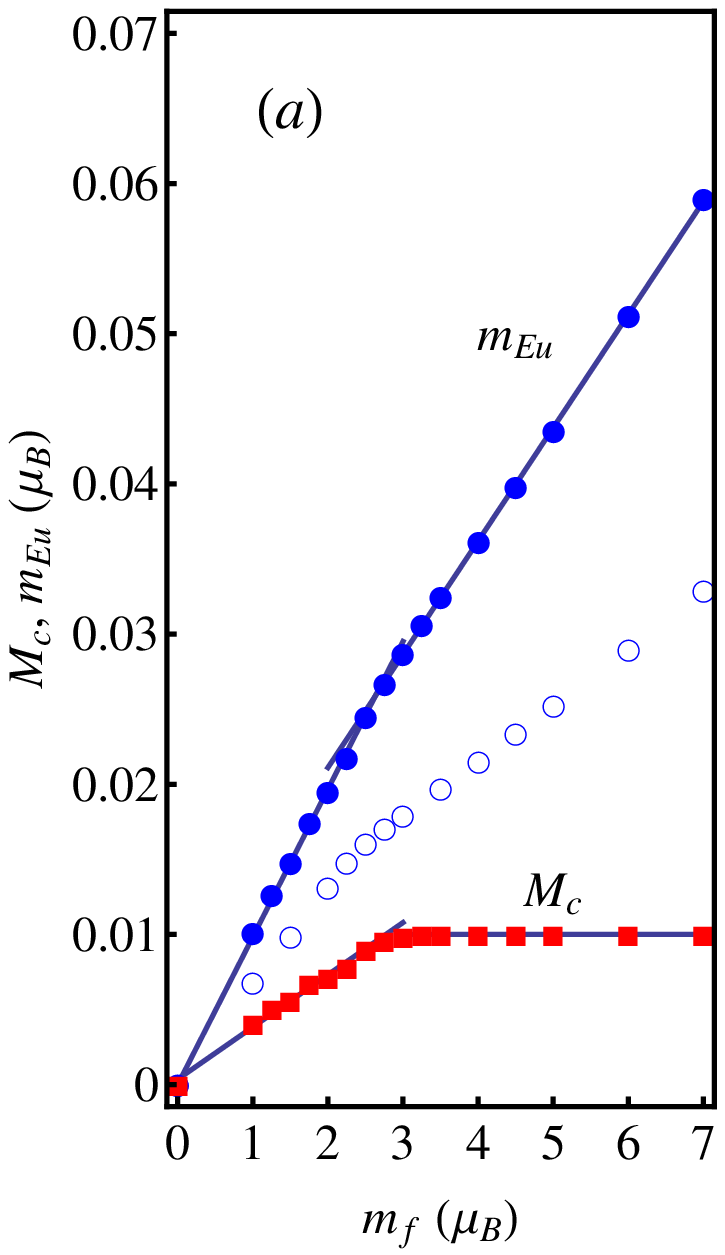}\hfil
\includegraphics[height=0.35\textwidth,angle=0,clip]{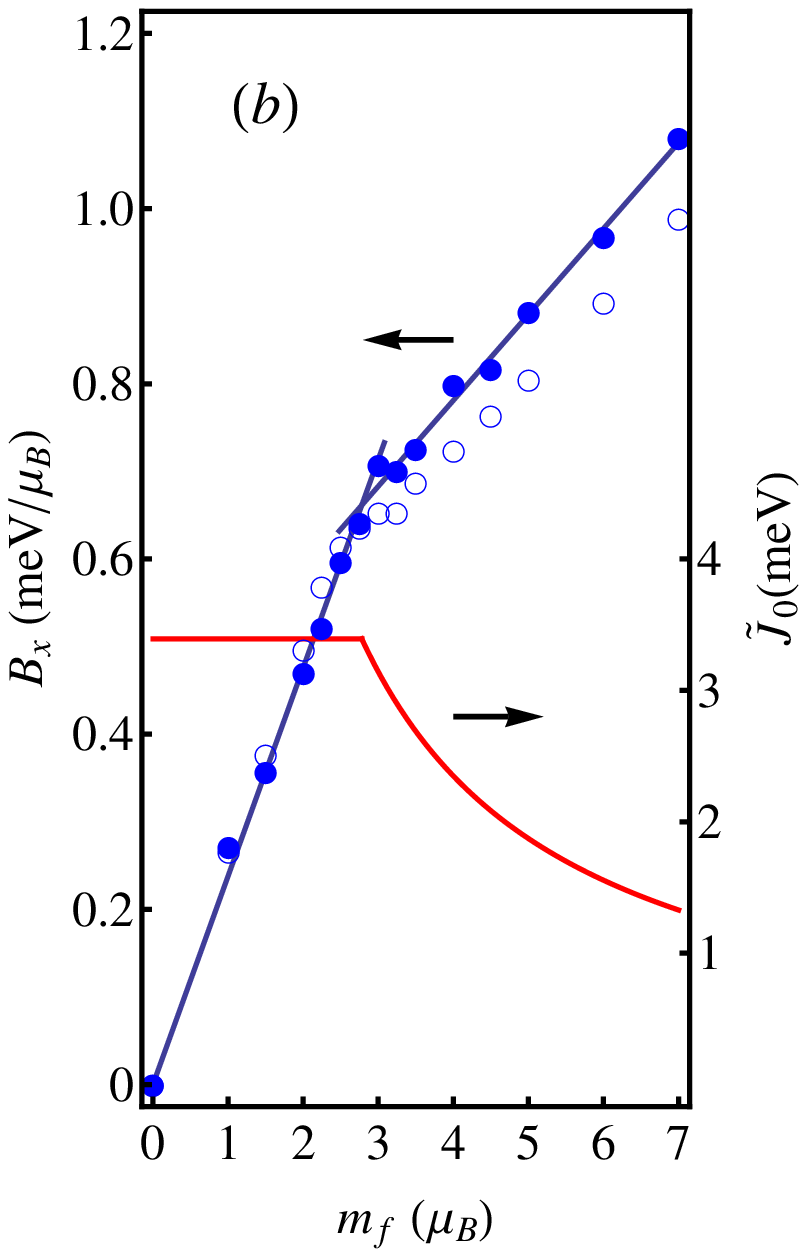}
\caption{(Color online) (a) Non-$4f$ contribution to the local moment on Eu atoms ($m_\mathrm{Eu}$, blue circles) and magnetization ($M_c$, red squares) as a function of the polarization $m_f$ of the $4f$ shell. (b) Effective exchange field on Eu ($B_x=J^\mathrm{Eu}_0/m_f$, blue circles) as a function of $m_f$ (left axis). In both panels the straight lines are linear fits to the data below and above the point of half-metallicity. All filled symbols are for Model 3, and empty symbols are for Model 2. Red solid line in panel (b): effective indirect exchange parameter $\tilde J^\mathrm{Eu}_0$ (right axis) extracted from the fits to $B_x$ for Model 3.}
\label{Var4f}
\end{figure}

So far we have assumed that the exchange interaction may be described by the Heisenberg model. However, at low doping concentrations the indirect exchange in EuO has a non-Heisenberg character due to the half-metallicity of the conduction band above the threshold polarization level. \cite{Mauger,Nagaev,Takahashi} In this half-metallic region the effective Weiss field no longer depends on the magnetization, which may explain the pronounced shoulder observed in the $M(T)$ curve (see e.\ g.\ Refs.\ \onlinecite{Sutarto,Mairoser} for recent data on high-quality samples). The effective exchange parameter in the paramagnetic state is larger compared to the ferromagnetic state, and thus the above estimates of $\Delta T_C$ should be corrected at low doping levels.
To this end, we have calculated the exchange fields as a function of the polarization of the $4f$ shell in EuO at 1\% and 3\% doping levels. These calculations were performed for Models 2 and 3 by varying the polarization of the open-core $4f$ shell, but this variation also represents the effects of temperature-dependent magnetization. The results for 1\% doping are displayed in Fig.\ \ref{Var4f}.

As seen in Fig.\ \ref{Var4f}a, the conduction-band magnetization $M_c$ increases almost linearly as a function of $m_f$ up to approximately $3\mu_B$. This is the point $m_f^c$ where the minority-spin conduction-band bottom moves above the Fermi level due to the exchange splitting induced by the $4f$ shell. For $m_f>m_f^c$ the magnetization stays constant at $0.01\mu_B$, which is equal to the electron doping level. However, the local moment on the Eu atom continues to grow at larger $m_f$ at a smaller rate. A similar change of slope is observed for the effective exchange field (Fig.\ \ref{Var4f}b). We interpret these features as a combination of indirect exchange interaction and other mechanisms. Indirect exchange contributes an approximately linear term to $m_\mathrm{Eu}$ and $B_x$ at $m_f<m^c_f$, but at $m_f>m^c_f$ its contribution stays constant. Other mechanisms (Bloembergen-Rowland) contribute a linear term for the entire range of $m_f$. Thus, the difference in slopes below and above $m^c_f$ gives the indirect exchange contribution to $B_x$, from which we can recover the effective indirect exchange $\tilde J^\mathrm{Eu}_0$. The latter is shown in Fig.\ \ref{Var4f}b by a solid (red) line. We see that $\tilde J^\mathrm{Eu}_0$ is 2.55 times larger in the paramagnetic state ($m_f\approx0$) compared to the ferromagnetic state ($m_f=7\mu_B$). This enhancement decreases with the doping level as the half-metallic region shrinks and eventually disappears. The corrected $\Delta T_C$ for 1\% and 3\% doping is shown in Fig.\ \ref{tcdop} by (red) triangles. Faster increase of $T_C$ with doping is in good agreement with experimental data.

\begin{figure}[htb]
\includegraphics[width=0.45\textwidth,angle=0,clip]{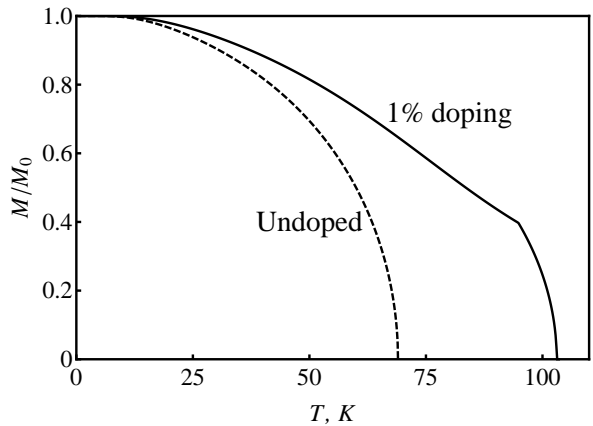}
\caption{Temperature dependence of reduced magnetization of EuO at 1\% doping taking into account the non-Heisenberg character of the indirect exchange interaction (solid line). The dashed line shows $M(T)$ for undoped EuO.}
\label{MofT}
\end{figure}

Following Mauger,\cite{Mauger} we can also estimate the effect of the non-Heisenberg character of the indirect exchange on the magnetization curve. Assuming that the electron gas remains degenerate, we solve the mean-field equations with magnetization-dependent indirect exchange contribution $\tilde J^\mathrm{Eu}_0$ added to the Heisenberg part. Instead of using the overestimated calculated value for the Heisenberg part of $J^\mathrm{Eu}_0$, we normalize it to produce the experimental $T_C$ of 69 K. The resulting $M(T)$ curve for 1\% doping is shown in Fig.\ \ref{MofT}. (Here we used the Brillouin function for quantum spin $7/2$, which makes the $T_C$ enhancement larger compared to that shown in Fig.\ \ref{tcdop}.) The shape of the $M(T)$ curve is similar to earlier calculations\cite{Mauger,Takahashi} and qualitatively similar to experiment (see e.\ g.\ Ref.\ \onlinecite{Sutarto}). The cusp, which is a consequence of our fitting procedure, should be smeared if a more accurate fitting is used near the crossover point $m^c_f$.

In this section we have considered the exchange interaction in electron-doped EuO within the virtual crystal approximation. In reality the individual dopants introduce local perturbations in the electronic structure and magnetic interaction due to their chemistry and structural relaxations around them. These effects are considered in the following sections.

\section{Electronic structure of G\lowercase{d} dopant and O vacancy in E\lowercase{u}O}\label{Isolated}

To gain more insight in the role of different electron dopants in EuO, in this section we consider isolated Gd impurities and O vacancies.
Here and in the next section we use the pseudopotential plane-wave method with the projected augmented wave (PAW) potentials \cite{Bloechl,VASP-PAW} as implemented in the VASP package.\cite{VASP}
Generalized gradient approximation (GGA) \cite{PBE} is employed in combination with the Hubbard $U$ correction for Eu and Gd $4f$ orbitals.
The atomic configurations were optimized in the ferromagnetic configuration using the conjugate-gradient algorithm. All the technical details of calculations were similar to Ref.\ \onlinecite{An-PRB}.

\subsection{Electronic structure in the ferromagnetic state}

\begin{figure}[htb]
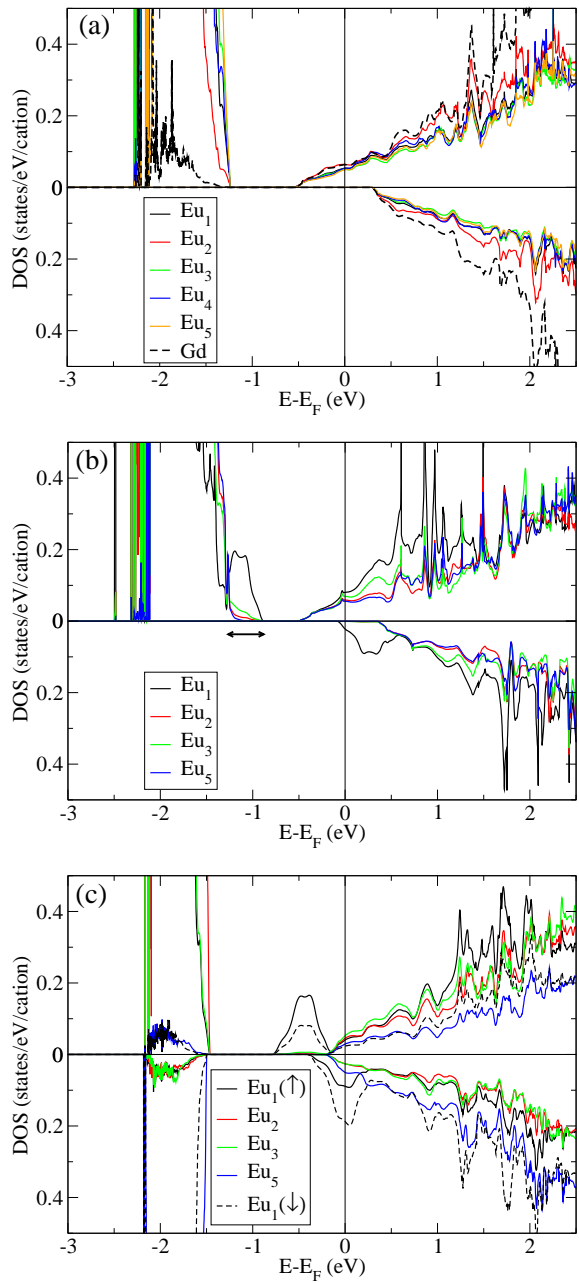

\includegraphics[width=0.42\textwidth,clip]{Fig6a.eps}\vskip3mm
\includegraphics[width=0.42\textwidth,clip]{Fig6b.eps}\vskip3mm
\includegraphics[width=0.42\textwidth,clip]{Fig6c.eps}
\caption{(Color online) Partial site-projected DOS for different cation sites.
(a) GdEu$_{31}$O$_{32}$ supercell, ferromagnetic state. (b) Eu$_{32}$O$_{31}$ supercell, ferromagnetic state. The arrow indicates the filled localized impurity state. (c) Eu$_{32}$O$_{31}$ supercell, random collinear spin configuration. The O vacancy has 3 Eu neighbors each with spin up and down, which are denoted Eu$_1(\uparrow)$ and Eu$_1(\downarrow)$. The cut-off peaks in all panels are the $4f$ states.}
\label{dos}
\end{figure}

Single impurities were introduced in the $2\times2\times2$ cubic supercell containing 64 atoms, producing GdEu$_{31}$O$_{32}$ and Eu$_{32}$O$_{31}$ structures.
As previously discussed, \cite{An-PRB} Gd impurities induce fairly large displacements of nearby O atoms towards themselves. Thus, in each nearest-neighbor Eu-O-Gd chain the Eu-O bond becomes 0.12-0.15 \AA\ longer than the Gd-O bond (see Fig.\ \ref{scheme}b for an illustration). Energy gained in this relaxation is responsible for the strong positive pair interaction for second-nearest neighbors in this system. In the case of an O vacancy, the neighboring Eu atoms shift 0.13 \AA\ away from the vacancy site (see Fig.\ \ref{scheme}c). Figure \ref{dos} shows site-projected densities of states (DOS) for the cations in the GdEu$_{31}$O$_{32}$ supercell (panel (a)) and in the
Eu$_{32}$O$_{31}$ supercell (panel (b)). The sites are labeled in the order of their distance from the impurity site. (For the supercell with an O vacancy, Eu$_4$ is equivalent to Eu$_1$.)

Each Gd atom donates one electron to the conduction band. As seen in Fig.\ \ref{dos}a, the deviations from homogeneous occupation of the conduction band are relatively small, indicating that the shallow impurity state is fully incorporated in the conduction band at the considered Gd concentration. Still, the occupied conduction-band partial DOS (Fig.\ \ref{dos}a) is somewhat larger for the Gd atom and its second-nearest Eu neighbors (Eu$_2$) compared to other Eu atoms. The increase for Gd is due to its larger nuclear charge, but for the Eu$_2$ atoms it is due to the displacement of the O atoms between Eu$_2$ and Gd towards Gd and away from Eu$_2$, which reduces charge transfer to O and shifts the electronic states on Eu$_2$ towards lower energies. The downward shift of the majority-spin $4f$ states on Eu$_2$ has the same origin. In the next section we will see that the associated changes in the local conduction-band magnetic moments significantly modify the exchange interaction near Gd impurities. Small differences in the conduction-band occupations between Gd and Eu atoms are in sharp disagreement with Ref.\ \onlinecite{Wang} arguing that Gd $5d$ states are occupied much more than the Eu $5d$ states. This difference might be due to the use of the Hubbard $U$ correction for the Eu $5d$ orbitals in Ref.\ \onlinecite{Wang}, which results in the band structure that disagrees with GW calculations.\cite{An-PRB}

\begin{figure}[htb]
\includegraphics[width=0.48\textwidth]{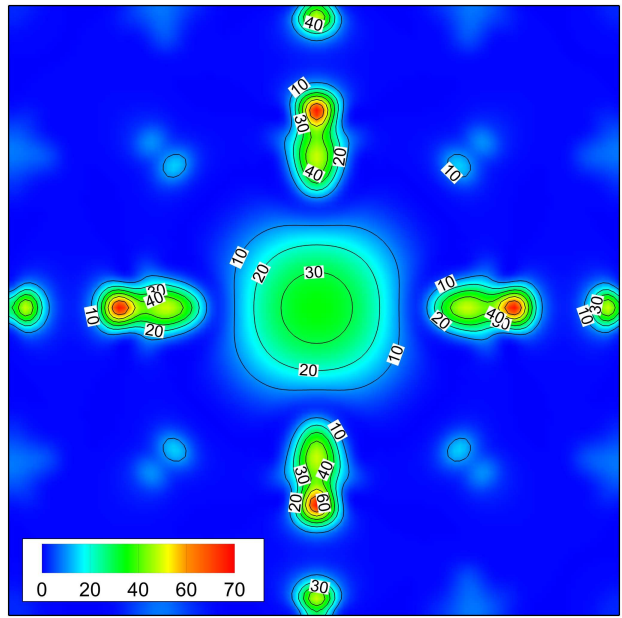}
\caption{(Color online) Charge density of the filled impurity state in the Eu$_{32}$O$_{31}$ supercell (units of 10$^{-3}$ \AA$^{-3}$).}
\label{CD}
\end{figure}

A charge-neutral O vacancy contributes two extra electrons. Fig.\ \ref{dos}b shows that in the ferromagnetic state only one of these electrons is donated to the conduction band, while the second one occupies a majority-spin impurity state at $-1.15$ eV, which does not extend beyond the Eu atoms adjacent to the vacancy (this state is indicated by the arrow in the figure). The minority-spin counterpart of this impurity state lies about 0.25 eV above the Fermi level. For the large vacancy concentration in our supercell, the impurity states are considerably broadened into impurity bands; as a result the minority-spin impurity band becomes slightly occupied. The energy dependence of DOS for the impurity states is typical for a single tight-binding band in the simple cubic lattice, which corresponds to the choice of the cubic supercell with one vacancy. The spatial structure of the occupied impurity state is seen from its charge density shown in Fig.\ \ref{CD}. The charge is concentrated on the vacancy site as well as on its six nearest Eu neighbors, extending further along the [001] directions towards the O atoms.

The exchange splitting of the impurity state on the O vacancy is approximately 1.4 eV, which is significantly larger than the exchange splitting of the empty conduction band in EuO. As will be further discussed below, a large part of this exchange splitting is due to the exchange-correlation field generated by the impurity state itself.

\subsection{Discussion of the thermally-induced effects in Eu-rich EuO}

The binding energy of the impurity states, their exchange splitting, and energy cost of double occupancy are important factors in the theories of the metal-insulator transition in Eu-rich EuO.\cite{Oliver,Oliver2,Torrance,Laks,Sinjukow} Let us, therefore, examine these features more closely. First we discuss the relevant parameters and then turn to their implications for the finite-temperature magnetic and transport properties.

\subsubsection{Temperature-dependent shifts}

Sinjukow and Nolting proposed a model \cite{Sinjukow} in which the impurity states in Eu-rich EuO are described in the atomic limit of the correlated Kondo lattice model.\cite{Nolting84} In their description each impurity state interacts with one Eu $4f$ spin. As a result, there are 4 quantum states for the ``impurity atom'' characterized by quantum numbers $n$ and $J$, where $n$ is the number of electrons in the impurity orbital, and $J$ is the total spin of the impurity atom: $|0,S\rangle$, $|1,S+1/2\rangle$, $|1,S-1/2\rangle$, and $|2,S\rangle$, where $S=7/2$ is the spin of the half-filled Eu $4f$ shell. The corresponding energies are 0, $\epsilon_0-JS/2$, $\epsilon_0+J(S+1)/2$, and $2\epsilon_0+U$, where $\epsilon_0$ is the binding energy of the impurity state in the absence of exchange coupling, $J$ is its exchange coupling with the $4f$ spin, and $U$ the Coulomb energy in the doubly occupied state. The four single-particle excitation energies correspond to the transitions $|0\rangle\to|1,S+1/2\rangle$, $|0\rangle\to|1,S-1/2\rangle$, $|1,S-1/2\rangle\to|2\rangle$, and $|1,S+1/2\rangle\to|2\rangle$, where the electron is extracted from the reservoir with the chemical potential $\mu$. The corresponding transition energies are $E_1=\epsilon_0-\mu-JS/2$, $E_2=\epsilon_0-\mu+J(S+1)/2$, $E_3=\epsilon_0-\mu+U-J(S+1)/2$, and $E_4=\epsilon_0-\mu+U+JS/2$. In this model the four transition energies do not depend on temperature, but their spectral weights do. \cite{Nolting84,Sinjukow} Based on these assumptions, it was argued \cite{Sinjukow} that the notion of a temperature-dependent exchange splitting of the impurity state \cite{Oliver2} is unphysical.

Contrary to the model assumption of Ref.\ \onlinecite{Sinjukow}, an impurity state on an O vacancy interacts not with one, but with six Eu $4f$ spins. Therefore, the atomic limit with four energy levels is irrelevant to this system, while the treatment based on the temperature-dependent exchange splitting is quite appropriate. Indeed, consider a cluster of six Eu spins interacting with an impurity spin and placed in an effective Weiss field $\mathbf{H}_W$ to describe the ferromagnetic state. This gives an effective Hamiltonian
\begin{equation}
\hat H_{eff} = -(\mathbf{H}_W+\tilde J\mathbf{\hat\sigma})\sum_{i=1}^6  \mathbf{\hat S}_i
\end{equation}
where $\mathbf{\hat\sigma}$ is the operator of the impurity spin. The empty and doubly occupied impurity states have spin zero and do not interact with the $4f$ spins at all. For the singly occupied impurity states $\mathbf{\hat\sigma}$ is the spin $1/2$ operator.
The singly occupied eigenstates can then be classified by the total spin $S_f$ of the $4f$ shells and its projection $S_{fz}$, plus the total spin of the system, as $|S_f+1/2,S_{fz}\rangle$ and $|S_f-1/2,S_{fz}\rangle$. The corresponding energy levels are $\epsilon_0-H_W S_{fz}-JS_f/2$ and $\epsilon_0-H_W S_{fz}+J(S_f+1)/2$. The states with 0 and 2 impurity electrons have energies $-H_W S_{fz}$ and $2\epsilon_0-H_W S_{fz}+U$.

The value $S_f$ lies within the range $0\leq S_f\leq 21$. Thus, there are multiple energy levels with different values of $S_{fz}$. This situation is quite different from the atomic limit, \cite{Sinjukow} in which the $4f$ spin is always equal to $7/2$. The statistical probabilities of levels with different $S_{fz}$ are determined by the Weiss field. At zero temperature only $S_{fz}=21$ is occupied, and the impurity levels are fully spin-split. The ground state is $|21+1/2,21\rangle$ and there are two single-particle transitions to empty and doubly occupied impurity orbital with energies $\epsilon_0-\mu-(21/2)\tilde J$ and $\epsilon_0-\mu+U+(21/2)\tilde J$. In the paramagnetic state, the most probable states have small values of $S_{fz}$, for which the exchange coupling with the impurity spin is weak. For these statistically important configurations the impurity state can be either singly or doubly occupied depending on the sign of $U+\epsilon_0-\mu$. Thus, up to statistical fluctuations around the mean-field solution, the situation is adequately captured by the temperature-dependent exchange splitting. In the following we will therefore disregard the unimportant quantum character of the $4f$ spins and treat them classically.

\subsubsection{O vacancy in the paramagnetic state}

In order to simulate the paramagnetic state, we repeated the calculation for the $2\times2\times2$ supercell containing one O vacancy, but now with a random assignment of spin directions (up or down) on all Eu atoms. The $4f$ moments add up to zero both in the entire cell and on the six Eu neighbors of the O vacancy. The self-consistent DOS shown in Fig.\ \ref{dos}c indicates that the impurity state in this configuration remains spin-split by $\Delta_x\approx 0.42$ eV. Since the $4f$ spins are randomized, this splitting is due to the exchange-correlation field intrinsic to the impurity state.

Fig.\ \ref{dos}c also shows that the impurity states have different weights on the Eu atoms with different orientations of the $4f$ spins. Specifically, an electron occupying a spin-up impurity state is about twice more likely to be found next to an Eu atom with spin up compared to an Eu atom with spin down. This feature indicates that the wave functions of the impurity states adjust to the given configuration of the $4f$ spins by partially localizing towards Eu atoms with parallel spins. The degree of this partial localization is controlled by the relation between the exchange splitting of the impurity state and the hybridization of the Eu $5d$ orbitals participating in its formation; it should not be sensitive to the vacancy concentration. Note that since the wave functions are different for different spin directions, the configuration of the doubly occupied impurity state is not a simple spin singlet.

\subsubsection{Estimates of the key parameters}

We will now estimate the parameters controlling the energies and occupation of the impurity states, including the binding energy $\epsilon_0$, the exchange couplings, and the charging energy $U$. We set the energy zero at the bottom of the unpolarized conduction band in the paramagnetic state. All estimates are intended to be accurate to no better than 0.05 eV.

\paragraph{Exchange energies.} By analyzing the local moments on different sites, we estimate that the impurity orbital has an approximately 38\% weight on the vacancy site itself, 34\% weight on the six nearest Eu sites, and the remaining weight extending to more distant sites. It is easy to see that the dominant part of $\Delta_x$ must be due to the exchange field generated by the ``head'' of the impurity orbital on the vacancy site. Indeed, the exchange-correlation kernel for the Eu $5d$ orbital is close to 1 eV, while the local moments on the nearest Eu sites are less than $0.06\mu_B$. This means that the contribution of Eu atoms to $\Delta_x$ is less than 0.02 eV.

The minority-spin impurity state is approximately half-filled. Therefore, the exchange-correlation energy for this orbital is $J_{imp}\approx 2\Delta_x\approx0.84$ eV. From the local moment and potential parameter splitting in an LMTO calculation we found that the exchange-correlation kernel $J_s$ for the $s$ orbital on the vacancy site is approximately 5 eV. This value is consistent with the magnitude of $J_{imp}$ ($J_{imp}\approx J_s\times0.38^2$).

In the paramagnetic state, the impurity levels are shifted downward by $\delta$ due to their partial localization. Since the weights on the Eu sites with parallel spins are about twice larger compared to those with opposite spins (Fig.\ \ref{dos}c), a simple arithmetic calculation shows that $\delta\approx\Delta_f/3$, where $\Delta_f$ is the downward shift of the majority-spin state in the ferromagnetic configuration due to exchange coupling with the $4f$ states.

The full exchange splitting of the impurity state in the ferromagnetic configuration is $\Delta_0=1.4$ eV. Since $\Delta_0=2\Delta_f+J_{imp}$, we find $\Delta_f\approx 0.28$ eV.

\paragraph{Charging energy $U_{imp}$.} In the ferromagnetic state (Fig.\ \ref{dos}b) the Fermi level lies close to zero energy (i.\ e.\ half-way between the majority and minority-spin CBM). The filled and empty impurity bands are centered approximately at $-1.15$ eV and 0.25 eV, respectively. The center of gravity of both bands is therefore at $-0.45$ eV. In the paramagnetic state (Fig.\ \ref{dos}c) the center of gravity of the impurity states is approximately at $-0.05$ eV (referenced from the CBM). The upward shift of $0.4$ eV is equal to $U_{imp}/2-\delta$ (the minority-spin impurity band is close to half-filling in Fig.\ \ref{dos}c). Thus, we find $U_{imp}\approx 0.98$ eV. This value is quite reasonable for an $F$ center-like state extending to nearest neighbors of the vacancy.

\paragraph{Binding energy $\epsilon_0$.} On the model level, the electron removal energy in the ferromagnetic state is equal to $-(\epsilon_0-\mu-\Delta_f)$. To estimate the removal energy from our DFT calculation, we need to correct the Kohn-Sham eigenvalue at $-1.15$ eV by subtracting $(U_{imp}-J_{imp})/2=0.07$ eV. Since $\mu\approx0$ in our ferromagnetic calculation, we find $\epsilon_0\approx -0.94$ eV.

\subsubsection{Occupation of the impurity states}

We are now in the position to discuss the temperature-dependent occupation of the impurity states. Let us first consider the limit of a very low vacancy concentration, for which the impurity bands do not broaden into bands, while the chemical potential lies close to the CBM. In the ferromagnetic state the CBM is exchange split, and $\mu\approx -0.4$ eV according to our calculations or $-0.3$ eV if we take the experimental value 0.6 eV for the conduction band splitting.\cite{Steeneken} The impurity orbital is obviously singly occupied by an electron with spin parallel to the magnetization. The excitation to an unoccupied state has energy $-(\epsilon_0-\mu-\Delta_f)\approx0.8$ eV, and the electron addition energy is $\epsilon_0-\mu+U+\Delta_f\approx0.7$ eV.

In the paramagnetic state we should set $\mu=0$ (CBM is not polarized), $\Delta_f=0$, and add the localization shift $-\delta$ to the impurity levels. The doubly occupied state now has energy $\epsilon_0-\mu+U-\delta\approx-0.05$ eV with respect to the singly occupied state, which means that the doubly occupied state becomes dominant. As long as there are charge carriers in the conduction band, it is preferable for them to get trapped by the impurity levels. We could arrive at the same result by following Slater's transition-state argument, which states that the addition energy can be estimated as a DFT eigenvalue half-way between the singly and doubly occupied configurations. In the random spin configuration (Fig.\ \ref{dos}c), the minority-spin impurity state happens to be very close to half-filling. Therefore, the addition energy can be estimated as the center of gravity of both impurity states minus the chemical potential. Since we are interested in the dilute limit, the chemical potential should be placed at the CBM.

Thus, our results suggest that slightly Eu-rich EuO should be semiconducting with an activation energy of about 0.05 eV. This picture is consistent with the so-called ``helium atom model'' of the metal-insulator transition in Eu-rich EuO,\cite{Oliver} in which the ferromagnetic order results in the change from double to single occupation of the impurity states and a subsequent promotion of free charge carriers into the conduction band. Eu-rich EuO samples undergoing the metal-insulator transition exhibit semiconducting conductivity in the paramagnetic state with an activation energy of about 0.3 eV.\cite{Oliver,Oliver2} Our estimate of 0.05 eV is therefore too small. The difference may be due to the finite size of the supercell, as well as to errors intrinsic to DFT. On the other hand, the observed activation energy may actually correspond to a different bound state. A bound magnetic polaron \cite{Torrance} is a popular candidate, in which the second localized electron aligns its spin parallel to the deeply bound impurity state and polarizes the spins of the Eu atoms in its vicinity. We note, however, that double occupation of $F$ center-like impurity states in the paramagnetic state is consistent with the experimental observation that $T_C$ in bulk Eu-rich EuO samples is not increased compared to its value for pure EuO, while even small concentrations of Gd enhance it notably. If bound magnetic polaron is the dominant defect in Eu-rich EuO, we would expect a $T_C$ enhancement of a magnitude similar to Gd doping at approximately double the concentration of O vacancies.

As we argued above, in the dilute limit O vacancies in EuO should capture two electrons in the paramagnetic state. The situation changes at larger vacancy concentrations on the order of a few percent, which can be reached in films. \cite{Massenet,Borukhovich,Matsumoto,Barbagallo} Compared to the dilute limit, there are two changes: the impurity levels broaden into impurity bands, and the chemical potential is increased unless there are compensating defects. Although the increase of the chemical potential tends to further stabilize the doubly occupied state, the lowest-energy state can now correspond to a fractional occupation of the impurity band. This is the situation seen in Fig.\ \ref{dos}c, where it corresponds to the vacancy concentration of 3.125\%. Of course, there should be no metal-insulator transition in this case, and $T_C$ should be significantly increased due both to the filling of the conduction band and to the exchange coupling mediated by the magnetically polarized impurity states, as is usually observed in Eu-rich EuO films.

In principle, the spin-polarized impurity orbitals could interact directly and order ferromagnetically above the $T_C$ of pure EuO. However, the following considerations show that this interaction is too weak.
In order to interact strongly, two O vacancies must share an Eu neighbor, which means they must occupy nearest or next-nearest O sites. The exchange coupling for such pairs can be estimated as the difference in the exchange-correlation energy $I_{5d} M^2/2$ ($I_{5d}\approx 1$ eV) on the shared Eu atom(s) for parallel and antiparallel orientations of the impurity spins. At a 3\% vacancy concentration, the vacancy orbitals carry local moments of about 0.5 $\mu_B$ (Fig.\ \ref{dos}c). Taking into account that the weight of the impurity orbital on each neighboring Eu atom is about 5.6\%, the local moment on Eu for parallel and antiparallel orientations should be about 0.056 $\mu_B$ and 0, respectively, yielding an energy difference of about 1.5 meV. If the vacancy orbitals are fully spin-polarized (local moment 1 $\mu_B$), this difference would be 4 times larger, or about 6 meV. At 3\% concentration each vacancy has a roughly 50\% chance of having another vacancy as its first or second neighbor. Thus, even in the mean-field approximation the maximum possible $T_C$ due to this exchange interaction is less than 10 K, which is well below the ordering temperature of pure EuO.

Thus, we expect that the vacancy orbitals should behave as Anderson local moments, and their dominant effect on $T_C$ should be through an enhancement of the exchange coupling between the Eu $4f$ spins that they overlap with. The local moment on the impurity orbital varies gradually in the whole range of magnetization from ferromagnetic to paramagnetic (see Fig.\ \ref{dos}b and \ref{dos}c), and the impurity band does not become half-metallic at any temperature. Therefore, it is likely that the interaction introduced by these orbitals does not deviate strongly from the Heisenberg model. The magnitude of this interaction will be evaluated in Section \ref{MSCE:vac}.

To conclude this section, we note that filling of the impurity states on O vacancies suggests a possible explanation for the results of the ARPES measurements on Gd-doped EuO, \cite{Shen} which showed the transfer of the spectral weight from the electron pockets near X points toward bound states of unknown origin as the system is heated beyond the Curie temperature. It was suggested \cite{Shen} that Gd-induced O vacancies could be responsible for a defect band contributing the bound states. Our present results not only provide evidence in support of this hypothesis, but also show the possible mechanism of the temperature-dependent spectral weight transfer. The Fermi level is shifted upwards in Gd-doped EuO, tending to stabilize the doubly occupied impurity state. Therefore, the impurity bands may get fully filled in the paramagnetic state even at fairly large vacancy concentrations.

\section{Magnetostructural cluster expansion}\label{sec:MSCE}

In this section we describe a magnetostructural cluster expansion (MSCE) approach, which is a combination of the cluster-expansion (CE) technique \cite{CW,Duc,Sanchez,deFon,Zunger,Walle} for substitutional alloys with the spin-cluster expansion \cite{Drautz} for magnetic systems, enabling the description of the interaction between the configurational and magnetic degrees of freedom in a magnetic alloy. This approach allows one to take into account the dependence of the magnetic interaction parameters on the local chemical environment, as well as the influence of magnetic order on the configurational thermodynamics. Our goal here is to evaluate the effects of Gd doping and O deficiency on the magnetic properties of EuO beyond the virtual crystal approximation, and in the actual calculations we will assume that the dopants are randomly distributed over the lattice. The main idea of this method and its application to Gd-doped EuO were briefly reported earlier.\cite{SPIE} Here we provide a more general formulation of the method and additional details justifying the choice of parameters for Gd-doped EuO. A simple MSCE model for O-deficient EuO is also presented.

In general, we may consider a magnetic substitutional alloy, whose configuration is described by occupations numbers $n_{ip}$, where $i$ is a lattice site, $p$ the component index, and $n_{ip}=1$ if and only if site $i$ is occupied by component $p$. Atoms of some components also have a classical spin variable $S_i$ associated with them, which we treat in the adiabatic approximation.\cite{Gyorffy} We assume that the total enthalpy of the alloy is a function of all occupation numbers and spin variables: $E=E(n,S)$. This total energy can be, quite generally, represented as an expansion in a set of spin-structure clusters
\begin{equation}
E(n,S)=\sum_{\alpha\subset A,\nu} m^\nu_{A\alpha} J^\nu_{A\alpha}\langle\Pi_A(n)\Phi^\nu_\alpha(S)\rangle
\label{msce}
\end{equation}
Here $A=\{i_1\to p_1,\ldots,i_N\to p_N\}$ denotes a subset of lattice sites (of size $N$) with a mapping to the set of components, and $\Pi_A$ is the corresponding projector: $\Pi_A=n_{i_1p_1}\cdots n_{i_Np_N}$. The index $\alpha$ designates a magnetic subset of $A$ (with the inherited mapping). The effective magnetostructural cluster interaction $J^\nu_{A\alpha}$ is an energy term attached to a particular spin basis function \cite{Drautz} $\Phi^\nu_\alpha$ (labeled by $\nu$) for a magnetic cluster $\alpha$. The angular brackets denote averaging over the entire crystal, and the factor $m^\nu_{A\alpha}$ is the multiplicity factor needed to count the clusters correctly depending on their symmetry.\cite{Walle} The influence of the alloy configuration on the magnetic interaction is reflected in the dependence of the parameters $J^\nu_{A\alpha}$ on $A$.

\subsection{Gd-doped EuO}

Although in general the expansion in Eq.\ (\ref{msce}) can include arbitrary multispin interactions, here we restrict ourselves to Heisenberg exchange. The resulting model is therefore not expected to correctly describe the non-Heisenberg indirect exchange at low doping levels, which was discussed in Section \ref{VCA}. The exchange parameters, however, will depend on the chemical environment. In this special case, the magnetic subcluster $\alpha$ always contains two sites, the index $\nu$ is superfluous, and the total energy can be written as
\begin{equation}
E=-\frac12\sum_{ij} J_{ij}(n)\mathbf{e}_{i}\mathbf{e}_{j}
\label{Heis}
\end{equation}
where $\mathbf{e}_{i}=\mathbf{S}_i/S_i$, and the dependence on the components residing at sites $i$ and $j$, as well as on the environment, is schematically absorbed in the configuration dependence of the exchange parameter $J_{ij}(n)$.

\begin{figure}[hbt]
\includegraphics[width=0.45\textwidth,clip]{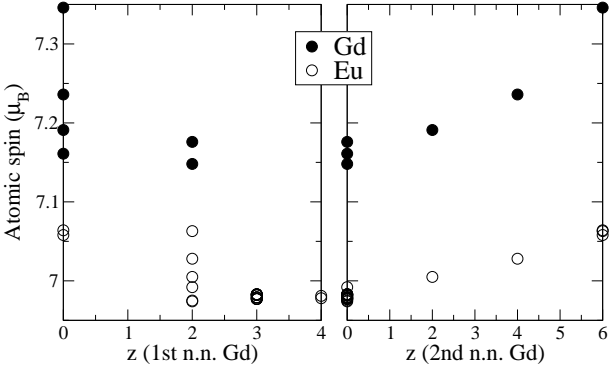}
\caption{Local magnetic moments of Gd (filled circles) and Eu (empty circles) in two Gd$_3$Eu$_{13}$O$_{16}$ structures as a function of the number of their nearest (left panel) and next-nearest (right panel) Gd neighbors.}
\label{M_of_z}
\end{figure}

\begin{figure}[hbt]
\includegraphics[width=0.45\textwidth,clip]{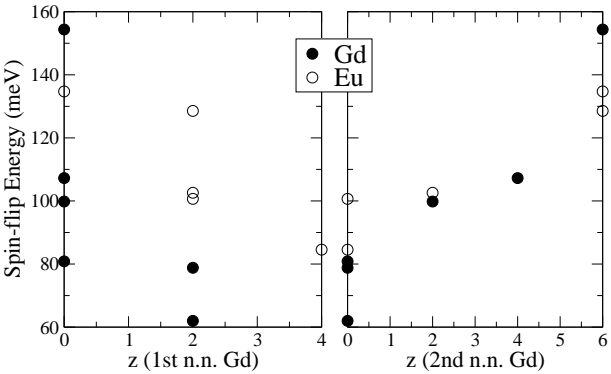}
\caption{Same as in Fig.\ \ref{M_of_z} but for spin-reversal energies instead of local moments.}
\label{E_of_z}
\end{figure}

As explained in Section \ref{Isolated}, the shifts of the O atoms toward Gd increase the local moments of Eu atoms that have Gd as their second-nearest neighbors. The same applies to Gd atoms that have next-nearest Gd neighbors. Fig.\ \ref{M_of_z} shows the dependence of Gd and Eu local moments in two representative Gd$_3$Eu$_{13}$O$_{16}$ structures on the number of first and second-nearest Gd neighbors. It is evident that the local moments strongly correlate with the number of next-nearest Gd neighbors, but not with the number of nearest Gd neighbors.
Further, Fig.\ \ref{M_of_z} shows the dependence of the energy needed to reverse an individual spin of Eu or Gd on the number of nearest or next-nearest Gd atoms. The same trend is observed as for the local moments in Fig.\ \ref{E_of_z}. Clearly, larger local moments indicate increased conduction-electron density and enhanced indirect exchange interaction.

\begin{figure}[hbt]
\includegraphics[width=0.4\textwidth]{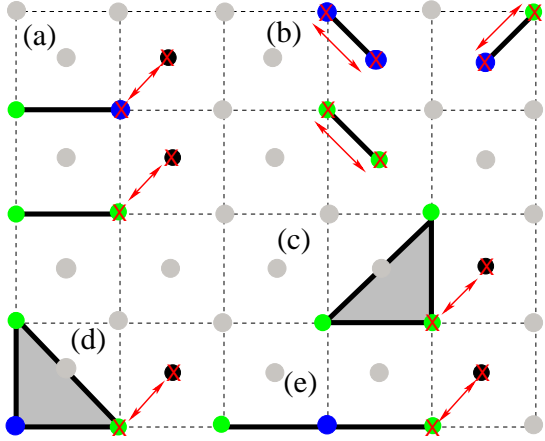}
\caption{(Color online) Clusters included in the MSCE (\ref{Jij})-(\ref{Si}). Red crosses connected by arrows: sites $i$, $j$.
Green circles: Gd atoms; blue circles: Eu atoms; black circles: either Gd or Eu. The clusters enter the MSCE with coefficients: (a) $B_p$; (b) $J^{pq}_{ij}$;
(c) $D$; (d) and (e) $C$.}
\label{graphH}
\end{figure}

This effect can be explained by the fact that some of the O neighbors of such Eu and Gd atoms are shifted away from them. As a result, their electronic states are lowered, and the conduction states become more populated. The increased local moments lead to enhanced exchange coupling with the neighboring cations. We used this physical insight to select the magnetostructural cluster set, using the standard cross-validation score \cite{Walle} as an indicator of the fit quality. This empirical selection resulted in the choice:
\begin{equation}
J_{ij}(n) = \sum_{pq}n_{ip}n_{jq}J^{pq}_{ij}(x)+\eta_{ij}[S_i(n)+S_j(n)],
\label{Jij}
\end{equation}
where the dependence of the exchange parameters on the concentration $x$ of Gd is included explicitly to represent the homogeneous effect of electron doping.
In the following, $p=e$ stands for Eu and $p=g$ for Gd. The parameters $J^{ee}_{ij}$ are non-zero for first and second neighbors; their values for $x=0$ and ratio $J^e_2/J^e_1$ for all $x$ are taken from the separate fitting for pure EuO;
and the dependence on $x$ is represented by multiplying $J^{ee}_{ij}(0)$ by a quadratic polynomial of $x$
with a unit free term.
The parameters $J^{eg}_{ij}$ are non-zero only for nearest neighbors and are represented by a quadratic polynomial. The parameters $J^{gg}_{ij}$ are non-zero for first and second neighbors and are independent of $x$. The second term in Eq.\ (\ref{Jij}) includes configurational corrections for nearest-neighbor exchange: $\eta_{ij}=1$ if $i$, $j$ are nearest neighbors and 0 otherwise. The configuration-dependent terms are defined as:
\begin{align}
S_i(n) &= \sum_p B_p n_{ip} Z_i + C n_{ig}\sum_{k\in M_i}(Z_k-1)n_{ke} \nonumber \\
&+ \frac12 D n_{ig} \sum_{kl\in M_i}\epsilon^i_{kl}n_{kg}n_{lg}
\label{Si}
\end{align}
where $B_e$, $B_g$, $C$, and $D$ are the fitting parameters, $M_i$ is the set of 6 second-nearest neighbors of site $i$, $Z_i=\sum_{k\in M_i}n_{kg}$ is the number of Gd atoms occupying the sites in $M_i$, and $\epsilon^i_{kl}=1$ only if the directions $i\to k$ and $i\to l$ are orthogonal (otherwise $\epsilon^i_{kl}=0$).
All the cluster types in Eqs.\ (\ref{Jij})-(\ref{Si}) are depicted in Fig.\ \ref{graphH}.

Note that the first term in Eq.\ (\ref{Jij}) corresponds to $\alpha=A$ in the general expansion of Eq.\ (\ref{msce}),
and its three nonequivalent clusters are shown separately in Fig.\ \ref{graphH}(b).
The remaining terms with $S_i$ given by Eq.\ (\ref{Si}) correspond to more complicated clusters $A$, $\alpha\subset A$.
The structure of these clusters is selected so as to incorporate the dominant effects of interactions mediated by the
local relaxations of the O atoms.
These interactions ``propagate'' between the next-nearest neighbor pairs, and a few clusters of this type are included
in Eq.\ (\ref{Si}).
In particular, the cluster corresponding to the last term in Eq.\ (\ref{Si}) is a right triangle of Gd atoms,
whose two sides connect next-nearest neighbors.
Each of the other two terms in Eq.\ (\ref{Si}), in fact, corresponds to more than one nonequivalent type of cluster.
The first term defines inequivalent clusters with different energies $B_p$, while the second term includes two inequivalent clusters with the same energy $C$.
This choice reduces the number of fitting parameters and is empirically justified by the quality of the fit.
Note that the subtraction of 1 in the second term of Eq.\ (\ref{Si}) is purely conventional and is designed to remove the explicit ``self-interaction.''

For the fitting of the MSCE (\ref{Jij})-(\ref{Si}) we used 41 supercells containing up to 64 atoms with different configurations of Gd dopants with concentrations up to 25\%. Full structural optimization was performed in the ferromagnetic spin configuration. Spin-reversal energies for a subset of individual Eu and Gd atoms in these supercells were then calculated. The structures used and the calculated spin-reversal energies are listed in the Supplementary Material. The resulting MSCE has a CV score of 6.6 meV/spin and the root-mean-square misfit of 5.6 meV/spin, and the fitted parameters are:
$J^{ee}_1(x)=0.962+12.9x-45.5x^2$; $J^{ee}_2/J^{ee}_1=0.02$; $J^{eg}_1(x)=0.293 + 13.5x -43.4x^2$; $J^{gg}_{1}=-0.605$; $J^{gg}_{2}=1.040$; $B_e=0.172$; $B_g=0.094$; $C=0.025$; $D=0.097$ (all in meV).
The parameters $B_e$, $B_g$, $C$, and $D$ correspond to 3-site and 4-site clusters of high multiplicity, and their impact on the quality of the fit is, thereby, quite significant. Note that the $x$-dependent contributions to both $J^{ee}_1$ and $J^{eg}_1$ are almost identical, reflecting the same mechanism of doping-induced interaction. The value of $J^{eg}_1$ is much smaller than $J^{ee}_1$ at $x=0$, which is reasonable, because the occupied Gd $4f$ orbitals are strongly bound and do not participate in the $f$-$d$ hopping mechanism. Antiferromagnetic $J^{gg}_1$ coupling is also reasonable in view of the fact that GdS is and GdO is predicted to be antiferromagnetic.\cite{An-PRB} Concentration-dependent coupling between two Gd atoms was not included, because it did not improve the quality of the fit.

The MSCE (\ref{Heis}) assumes the Heisenberg form of magnetic interaction and is fitted to configurations that differ from ferromagnetic by one spin reversal. As we discussed in Section \ref{VCA}, at small doping levels the indirect exchange has a non-Heisenberg character, because the conduction band becomes half-metallic at sufficiently large magnetization. Therefore, the MSCE is expected to underestimate $T_C$ at small doping levels. Still, we found that the total energy differences between randomly selected magnetic configurations of the GdEu$_{31}$O$_{32}$ supercell are faithfully reproduced by the Heisenberg MSCE, even though their average energy is underestimated by MSCE by about 13\%.\cite{SPIE} Therefore, we may expect that deviations from Heisenberg behavior have a relatively small effect on $T_C$ for concentrations of 3\% or more. This is in agreement with VCA results shown in Fig.\ \ref{tcdop}.

\subsection{O-deficient EuO}
\label{MSCE:vac}

For O-deficient EuO, we would like to understand the large enhancement of $T_C$ observed in Eu-rich EuO films, which may contain much higher vacancy concentrations.\cite{Massenet,Borukhovich,Matsumoto,Barbagallo} Such films may be strongly heterogeneous, \cite{Borukhovich,Altendorf} although there is also evidence against phase separation.\cite{Monteiro} In order to compare strongly O-deficient with Gd-doped EuO, we constructed a simple MSCE for the vacancy concentration of 3.125\% under an assumption that the configurations with vacancies appearing next to each other are statistically insignificant. Thus, we considered one Eu$_{32}$O$_{31}$ supercell with one O vacancy (same as in Section \ref{Isolated}) and calculated the total energies of several magnetic configurations with reversed spins on Eu atoms. As in the MSCE for Gd-doped EuO, we assume that the magnetic interaction has a Heisenberg form, which is likely a fair approximation for the 3\% electron doping level (see the discussion in Sections \ref{VCA}, \ref{Isolated} and a further comment below).

There are four inequivalent Eu atoms in the chosen supercell. The energy required to reverse the spins on the Eu atom adjacent to the vacancy is 119.5 meV, while those for the more distant Eu sites range from 63 to 69 meV. Since these latter energies are very similar, we assumed that the exchange interaction is homogeneous beyond the nearest neighbors of the vacancy and used the average value 67.6 meV for it. This value includes the contribution of homogeneous electron doping. Large enhancement of the spin-reversal energy for the nearest neighbor of the O vacancy is due to the impurity state (Section \ref{Isolated}) mediating strong ferromagnetic coupling between the Eu atoms with which it overlaps. This is the main feature that we wish to include in the MSCE. In order to resolve this interaction by pairs of atoms, we also calculated the energy changes needed to reverse two of the six Eu neighbors of the vacancy. There are two inequivalent configurations of this kind, and the corresponding energies are 221.2 meV (when two Eu atoms are nearest neighbors) and 234.4 meV (when they are not).

The calculated total energies can be well fitted using Eq.\ (\ref{Heis}) with
\begin{equation}
J_{ij}(n) = A J^{0}_{ij} + \eta_i \eta_j (S_1\eta_{ij} + S_2\tilde\eta_{ij})
\label{smvac}
\end{equation}
where $J^0_{ij}$ is the interaction determined separately for pure EuO, $A$ is the scaling factor accounting for the effect of homogeneous electron doping, $\eta_i=1$ if site $i$ has a vacancy at an adjacent O site, $\eta_{ij}=1$ if $i$, $j$ are nearest neighbors (same as in Eq.\ (\ref{Jij})), and $\tilde\eta_{ij}=1$ if $i$, $j$ are second-nearest neighbors.
The fitted parameters 
are $A$=1.45, $S_1=3.08$ meV and $S_2=2.43$ meV. The root-mean-squared error for the predicted magnetic energies is 4.1 meV.
This representation is designed for a particular vacancy concentration (1 per 32 O sites).

To quantify the deviations from Heisenberg behavior, we have compared the prediction of the MSCE (\ref{smvac}) for the Eu$_{32}$O$_{31}$ supercell with randomized directions (its DOS is shown in Fig.\ \ref{dos}c) with the calculated total energy. The predicted total energy difference with respect to the ferromagnetic state is underestimated by about 11\%, which is similar to the Gd-doped system at the same concentration. This underestimation can again be attributed to the half-metallicity of the conduction band in the ferromagnetic state, and its small magnitude supports the use of the Heisenberg model at this vacancy concentration.

The parameter $A$ in (\ref{smvac}) represents the homogeneous enhancement of the exchange interaction (relative to pure EuO) due to the introduction of free charge carriers, and the same role is played by the $x$-dependent $J^{ee}_1$ parameter in the case of Gd doping. As we've seen in Section \ref{Isolated}, both Gd and O vacancy donate one electron to the conduction band. Using the fitted expression for $J^{ee}_1(x)$ for Gd-doped EuO from Section \ref{sec:MSCE}, we find that at $x=1/32$ it is enhanced by a factor 1.37 compared to $x=0$. This value is similar to the parameter $A$ for Eu-rich EuO. On physical grounds we may expect that in a reasonable range of O deficiency the parameter $A$ should change similar to $J^{ee}_1(x)$ in Gd-doped EuO, while the other parameters ($S_1$ and $S_2$) reflecting the contribution from the localized state should be independent on the concentration.

We have also considered the possibility of the formation of complex defects when both Gd substitution and O vacancies are simultaneously present. To this end, we have calculated the total energies of a GdEu$_{31}$O$_{31}$ supercell, which contains one substitutional Gd and one O vacancy, for their 4 different mutual arrangements in this $2\times2\times2$ supercell. For these arrangements the Gd and vacancy sites are connected by vectors $(1/2,0,0)$, $(1/2,1/2,1/2)$, $(1,1/2,0)$, and $(1,1,1/2)$ in units of the lattice parameter $a$. The lowest total energy is achieved for the second of these arrangements, relative to which the four configurations are 138, 0, 46, and 19 meV higher, respectively. As Gd-doped EuO films are usually grown at temperatures of 350$^\circ$C or higher,\cite{Sutarto,Mairoser} these results suggest that O vacancies should occupy the sites next to Gd atoms less often than in the random alloy, but there is no preference for the formation of complex defects.

\section{Thermodynamic properties}\label{sec:MC}

The thermodynamic properties of Gd-doped and Eu-rich EuO were studied using classical Monte Carlo simulations for the MSCE models constructed in Section \ref{sec:MSCE} and the Metropolis algorithm.
The simulation box contained up to 32000 cation atoms, and the dopants were randomly distributed over their sublattice.  The equilibration and averaging passes for each temperature were performed with
$5\times10^5$ Monte Carlo steps per site or more.

\begin{figure}[htb]
\includegraphics[width=0.45\textwidth]{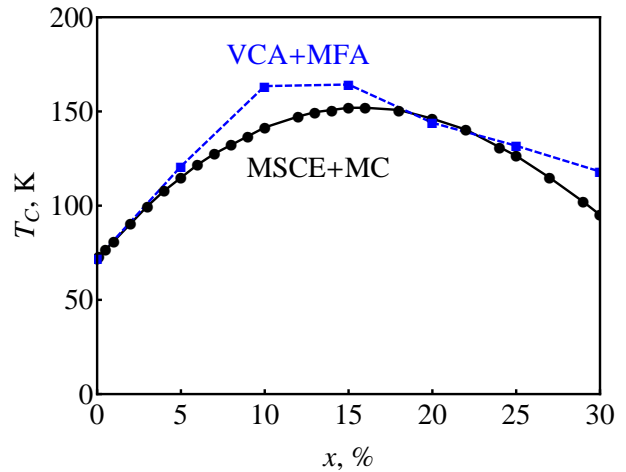}
\caption{(Color online) Curie temperature $T_C$ in Gd-doped EuO from Monte Carlo simulations for the MSCE model (black circles) and from the mean-field estimate based on linear-response results in VCA (blue squares). The latter is shifted by a constant so that $T_C(0)$ coincides with the MC result.}
\label{MC}
\end{figure}

The $M(T)$ curves for a wide range of Gd compositions in Gd$_x$Eu$_{1-x}$O do not reveal any unusual features.\cite{SPIE} Fig. \ref{MC} shows the Curie temperature $T_C(x)$ obtained from the inflection point of the $M(T)$ curve. For comparison, the plot of $\Delta T_C$ obtained in VCA for Model 1 (Fig.\ \ref{tcdop}) is added to the Monte Carlo value of $T_C$ at $x=0$ for the MSCE model. The Curie temperature for pure EuO is approximately 72 K in excellent agreement with experiment. The concentration dependence of $T_C$ is similar in VCA and in the MSCE-based MC. Thus, the effects of lattice contraction and local environment on the exchange interaction do not appear to modify the doping dependence strongly even at larger doping levels. The maximum $T_C$ of approximately 150 K is reached in a broad range of Gd concentrations in good agreement with experiments (see e.\ g.\ Ref.\ \onlinecite{Mairoser}). It is therefore unlikely that $T_C$ can be further increased in Gd-doped EuO. Similar to VCA, the MSCE-based MC results should underestimate the rate of increase of $T_C$ at low doping levels due to the neglect of the non-Heisenberg character of the indirect exchange.

\begin{figure}[hbt]
\centering
\includegraphics[width=0.45\textwidth]{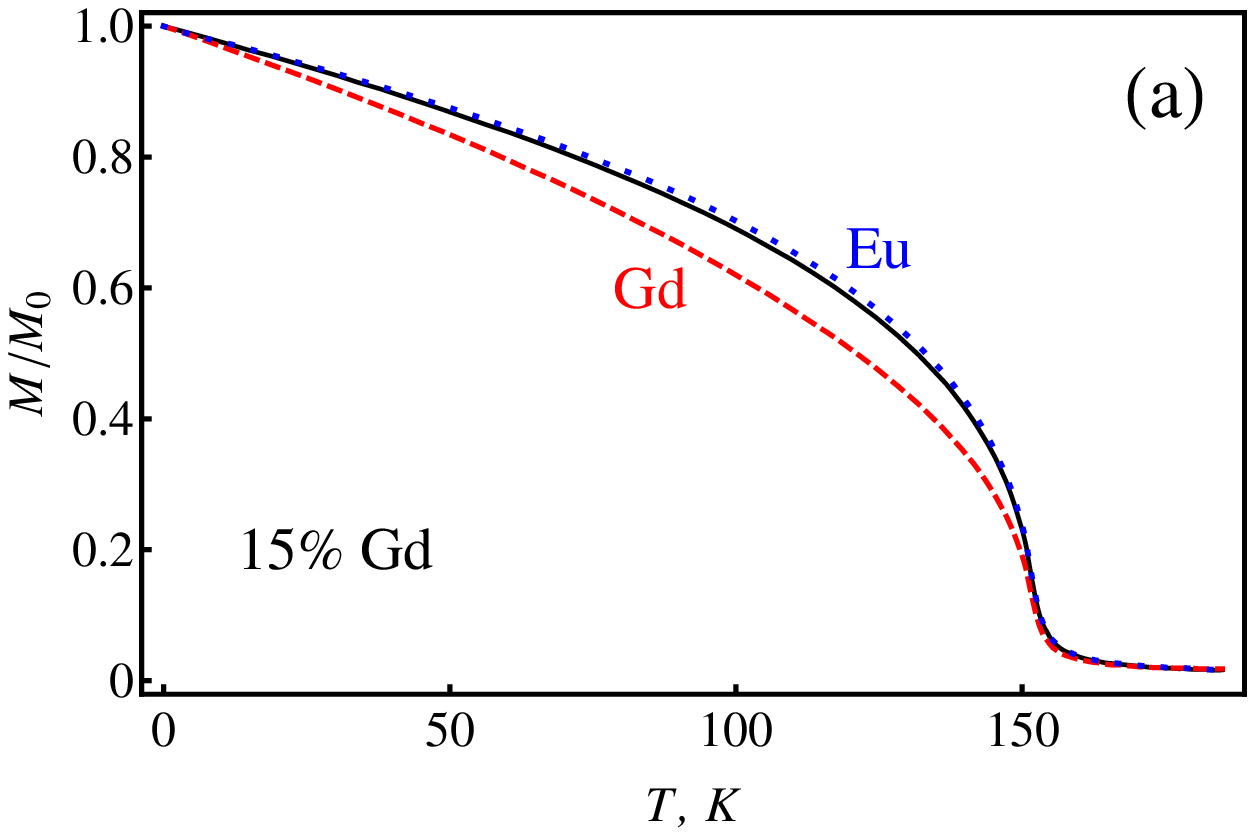}
\includegraphics[width=0.45\textwidth]{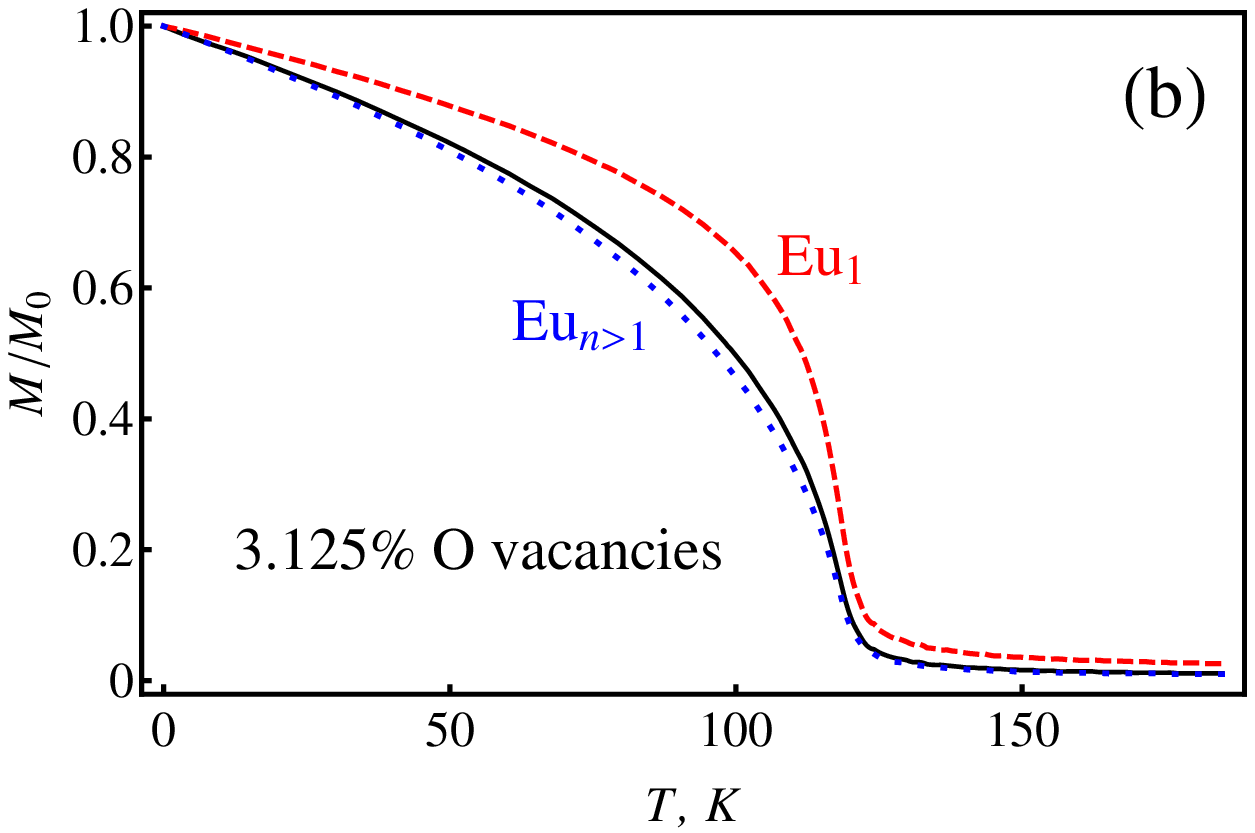}
\caption{(Color online) Atom-resolved $M(T)$ curves in (a) 15\% Gd-doped, and (b) 3.125\% Eu-rich EuO. Solid lines show the total magnetization. In panel (b) the atomic magnetization of the Eu neighbors of O vacancies (denoted Eu$_1$) is displayed separately from that of all other Eu atoms (denoted Eu$_{n>1}$).}
\label{MCD}
\end{figure}

In Gd-doped EuO there are two types of magnetic atoms, while in Eu-rich EuO the Eu neighbors of an O vacancy are strongly coupled by the localized state. It is therefore interesting to examine the contributions to the total magnetization coming from these different types of atoms. Fig.\ \ref{MCD}a shows the atomic contributions to the total $M(T)$ in 15\% Gd-doped EuO for Eu and Gd atoms. Fig.\ \ref{MCD}b provides a similar plot for Eu-rich EuO with a 1/32 vacancy concentration, sorting out the atomic contribution to $M(T)$ of Eu atoms that have an O vacancy next to them.

For Gd-doped EuO we see that the reduced magnetization of the Gd atoms is notably suppressed compared to that of the Eu atoms. This is because the Gd-Eu exchange coupling is significantly weaker compared to the Eu-Eu coupling, as mentioned in Section \ref{sec:MSCE}. The difference in the $M(T)$ curves for Eu and Gd atoms could be experimentally verified using an element-specific probe such as nuclear magnetic resonance.
As seen in Fig.\ \ref{MCD}b, in Eu-rich EuO there is a large enhancement of the magnetization for Eu sites adjacent to O vacancies, which is due to strong exchange coupling mediated by the localized states centered at the vacancies. The $T_C$ for the chosen vacancy concentration of 3.125\% is close to 120 K, which is only achieved at about twice larger Gd concentration in Gd-doped EuO. Thus, the homogeneous enhancement of the exchange interaction due to the filling of the conduction band and inhomogeneous coupling due to localized states at the vacancy sites contribute nearly equally to the enhancement of $T_C$ in Eu-rich EuO.

\begin{figure}[hbt]
\includegraphics[width=0.45\textwidth]{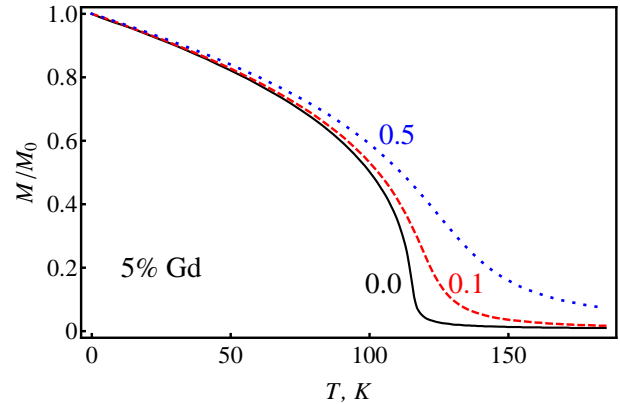}
\caption{(Color online) $M(T)$ curves for Gd$_{0.05}$Eu$_{0.95}$O in an external magnetic field. The field is indicated next to the curves in meV units of maximum Zeeman energy per site. For $7\mu_B$ magnetic moment, 1 meV is equivalent to $B=2.47$ T.}
\label{zeeman}
\end{figure}

Several authors observed that the deviations of the $M(T)$ curves from the Brillouin function with a long ``tail'' extending to higher temperatures is likely due, at least in part, due to the finite magnetic field used in the magnetization measurements.\cite{Altendorf,Burg,Mairoser2013} This feature is readily reproduced in MC simulations with a Zeeman field added to MSCE, as shown in Fig.\ \ref{zeeman}.

Overall, our results show that, although on the microscopic level the magnetic interaction in Gd-doped and Eu-rich EuO deviates strongly from the homogeneous electron doping model (VCA), this inhomogeneity does not strongly modify the macroscopic magnetic properties such as $T_C(x)$ and $M(T)$. The observed deviations of the $M(T)$ curves from the conventional behavior should therefore be due to other mechanisms. At low concentrations, as noted above, these deviations may be explained by the non-Heisenberg character of the indirect exchange in the half-metallic regime.\cite{Mauger,Nagaev,Takahashi} On the other hand, the observed anomalies at large Gd concentrations may be induced by finite external fields used in the measurements,\cite{Mairoser2013} or may indicate phase separation as suggested in Ref.\ \onlinecite{Sutarto}. The magnetization curves in Eu-rich EuO films usually look like a superposition of two separate magnetization curves with different Curie temperature, one of which coincides with that of pure EuO. This feature suggests phase separation,\cite{Borukhovich,Altendorf} but contrary evidence \cite{Monteiro} shows the need for further investigation.

\section{Conclusions}\label{Conclusions}

We have studied the electronic structure, magnetic interaction, and thermodynamic properties of Gd-doped and Eu-rich EuO using first-principles calculations. Homogeneous electron doping of EuO, which was modeled in the virtual crystal approximation, leads to a broad maximum in $T_C$. This maximum is due to the competition between the monotonically increasing indirect exchange contribution and the $f$-$d$ hopping mechanism, which generates ferromagnetism in pure EuO but is suppressed at large doping concentrations. Calculations with variable polarization of the $4f$ shell reveal a pronounced non-Heisenberg character of the indirect exchange interaction at small doping concentrations in agreement with the model of Mauger,\cite{Mauger} which leads to strongly enhanced $T_C$ and an anomaly in the $M(T)$ curve.

We have analyzed the electronic structure of isolated Gd impurities and O vacancies in EuO. Both these defects donate one electron to the conduction band in the ferromagnetic state. An O vacancy introduces an exchange-split $F$ center-like level, which is half-filled in the ground state. Estimates based on the key parameters extracted from our calculations suggest that in the limit of low vacancy concentrations the impurity state should become doubly occupied, assuming that the $4f$ spins are fully disordered. This is consistent with the ``He atom model'' of the metal-insulator transition observed in the bulk samples of Eu-rich EuO \cite{Oliver} and with no enhancement of $T_C$ in such samples. However, the estimated 0.05 eV activation energy in the paramagnetic state is too low compared to the transport measurements, and we can not rule out the formation of bound magnetic polarons. At higher vacancy concentrations typical for Eu-rich EuO films, the impurity states broaden into impurity bands, the upper one of which remains partially filled in the paramagnetic state. The impurity states behave as Anderson local moments in this regime. This is consistent with the observed strong enhancement of $T_C$ in Eu-rich EuO films. Localized states on O vacancies could also provide the bound states observed in ARPES in Gd-doped EuO. \cite{Shen}

To go beyond the homogeneous doping model, we employed a magnetostructural cluster expansion (MSCE) approach, which explicitly includes the effect of local chemical environment on the magnetic interaction. For the case of Gd-doped EuO we constructed an MSCE based on the input data from first-principles calculations that covers a wide range of Gd concentration of up to 25\%. For Eu-rich EuO, we set up a simple MSCE based on the data for one supercell with a single vacancy. The resulting Hamiltonians were used in Monte Carlo simulations. The Curie temperature in Gd-doped EuO was found to behave similar to the virtual crystal approximation, exhibiting a maximum as a function of Gd concentration. The results suggest that the intrinsic limit \cite{Mairoser} of $T_C$ in Gd-doped EuO is approximately 150 K. Oxygen vacancies induce a sharper rise in $T_C$ compared to a similar concentration of Gd dopants due to the strong exchange coupling of the neighboring Eu atoms mediated by the polarized impurity levels. Microscopic inhomogeneity does not lead to any anomalies in the $M(T)$ curves, at least in the random alloys that we have considered.

\acknowledgments

We thank S. V. Barabash, P. A. Dowben, and A. G. Petukhov for useful discussions. This work was supported by DTRA (Grant No. HDTRA1-07-1-0008) and NSF (Grant No.\ DMR-1005642). Computations were performed utilizing the Holland Computing Center at the University of Nebraska. K.\ D.\ B.\ acknowledges support from the Research Corporation through the Cottrell Scholar Award.

\end{document}